\documentstyle[12pt,aaspp4]{article}

\begin{document}
\noindent\bf Accepted for publication in The Astronomical Journal
\rm

\title{An IR--Selected Galaxy Cluster at $z=1.27$\altaffilmark{1}}

\altaffiltext{1}{Based in part on observations obtained at the W.M.\ Keck
Observatory}

\author{S.A.\ Stanford} 
\affil{Institute of
Geophysics and Planetary Physics, Lawrence Livermore National
Laboratory, Livermore, CA 94550} 
\authoremail{adam@igpp.llnl.gov}
 
\author{Richard Elston\altaffilmark{2}} \affil{Department of
Astronomy, University of Florida, Gainesville, FL 32611 }
\authoremail{prme@kromos.jpl.nasa.gov}

\author{Peter R.\ Eisenhardt\altaffilmark{2}} \affil{Jet Propulsion
Laboratory, California Institute of Technology, Pasadena, CA 91109}
\authoremail{prme@kromos.jpl.nasa.gov}

\altaffiltext{2}{Visiting Astronomer, Kitt Peak National
Observatory, National Optical Astronomy Observatories, which is
operated by the Association of Universities for Research in Astronomy,
Inc., under cooperative agreement with the National Science
Foundation.}

\author{Hyron Spinrad, Daniel Stern}
\affil{Astronomy Department, University of California, 
Berkeley, CA  94720}
\authoremail{spinrad@bigz.berkeley.edu, dan@bigz.berkeley.edu}

\and
\author{Arjun Dey}
\affil{National Optical Astronomy Observatories\altaffilmark{3}, 
Tucson, AZ  85726}
\authoremail{dey@noao.edu}

\altaffiltext{3}{The National Optical Astronomy Observatories are
operated by the Association of Universities for Research in Astronomy,
Inc., under cooperative agreement with the National Science
Foundation.}

\begin{abstract}
We report the discovery of a galaxy cluster at $z = 1.27$.  ClG J0848+4453
was found in a near--IR field survey as a high density region of objects with
very red $J-K$ colors.  Optical spectroscopy of a limited number of $24
\lesssim R \lesssim 25$ objects in the area shows that 6 galaxies within a 90
arcsec (0.49 $h_{100}^{-1}$ Mpc, q$_0=0.1$) diameter region lie at $z = 1.273
\pm 0.002$.  Most of these 6 member galaxies have broad--band colors
consistent with the expected spectral energy distribution of a
passively--evolving elliptical galaxy formed at high redshift.  An additional
2 galaxies located $\sim$2 arcmin from the cluster center are also at $z =
1.27$.  Using all 8 of these spectroscopic members, we estimate the velocity
dispersion $\sigma = 700 \pm 180$ km s$^{-1}$, which is similar to that of
Abell richness class R $=$ 1 clusters in the present epoch.  A deep
Rosat/PSPC archival observation detects X--ray emission at the 5~$\sigma$ level
coincident with the nominal cluster center.  Assuming that the X--ray flux is
emitted by hot gas trapped in the potential well of a collapsed system (no
AGN are known to exist in the area), the resulting X--ray luminosity in the
rest frame 0.1--2.4 keV band of $L_x = 1.5 \times 10^{44}$ ergs s$^{-1}$
suggests the presence of a moderately large mass.  ClG J0848+4453 is the
highest redshift cluster found without targetting a central active galaxy.
    
\end{abstract}


\section{Introduction}

The existence of massive collapsed objects, such as rich galaxy clusters, at
high redshift provides a challenge for theories of cosmic structure formation
(Press \& Schechter 1974).  Numerical simulations based on hierarchical
clustering models such as cold dark matter (CDM) offer useful predictions of
the evolution of the cluster number density with cosmic time ($e.g.$ Cen \&
Ostriker 1994a; Eke, Cole, \& Frenk 1996; Viana \& Liddle 1996; Bahcall, Fan,
\& Cen 1997).  The rate at which clusters form depends strongly on
$\Omega_0$, and weakly on $\Lambda$ and the initial power spectrum.  CDM
predicts that the cluster population evolves rapidly for $\Omega_0 = 1$, and
very slowly for a low--density universe (e.g.\ Peebles, Daly, \& Juszkiewicz
1989; Evrard 1989).  So far a census of high redshift clusters sufficiently
complete for such model tests has been difficult to assemble.

In addition to their utility in large--scale structure studies, high redshift
galaxy clusters provide important tools in the study of galaxy formation and
evolution.  The galaxy populations of rich cluster cores up to at least $z
\sim 1$ tend to be dominated by massive ellipticals, which are good probes of
galaxy evolution because their stellar populations appear to be relatively
simple.  Ellipticals in present--epoch clusters form a homogeneous population
(Bower, Lacey, \& Ellis 1992; De Propris et al.\ 1997), which has been largely
quiescent since at least $z \sim 1$ (Ara\'gon-Salamanca et al.\ 1993; Lubin
1996; Ellis et al.\ 1997; Stanford, Eisenhardt, \& Dickinson 1998, hereinafter
SED98).  The nature of elliptical galaxy formation at even higher redshifts
probably depends strongly on the importance of merging in assembling the
stellar mass.  Beyond $z\sim1$ in cosmologically flat CDM models, the amount
of merging occurring within the prior $\sim$1 Gyr is large and could seriously
inflate the locus of early--type galaxy colors due to interaction--induced
starbursts (Kauffmann 1996).  On the other hand, models in which ellipticals
formed by a monolithic collapse at high-$z$ (e.g.\ Eggen, Lynden-Bell, \&
Sandage 1962) predict a tight color--magnitude relation out to at least $z
\sim 2$ for reasonable cosmologies.  Thus, the identification of clusters at
$z > 1$ and the characterization of their galaxy populations provides a
powerful means of testing elliptical galaxy formation theories.

Finding high redshift clusters is difficult.  The field galaxy population
overwhelms traditional searches for two--dimensional overdensities in the
optical.  Most searches have targeted high redshift radio galaxies and
quasars under the assumption that such massive objects may be signposts of
early collapse and the existence of massive dark matter halos.  Several
probable $z > 1$ clusters have been found using this method, including 3C~324
at $z = 1.20$ (Dickinson 1995) and 53W002 at $z = 2.39$ (Pascarelle et al.\
1996).  Many other more tentative candidates have been published as well, in
which claims of cluster identification often are based only on $\lesssim$3
common redshifts, including the AGN.  3C~324 is the highest redshift example
of a massive cluster which has strong evidence of both a large number of
known members, based on extensive optical spectroscopy, {\it and} of a deep
potential well, in the form of extended X--ray emission (Dickinson 1996).

An emerging method of finding $z > 1$ cluster candidates is deep near--IR sky
surveys.  Because their near--IR light is dominated by the light of evolved
giant stars, the $J-K$ colors of nearly all galaxies are simple functions of
redshift out to $z \sim 2$.  Though still constrained by the relatively small
size of the available arrays, near--IR imaging surveys soon may offer a
viable alternative to X--ray searches, which at present are limited to the
detection of only the most massive clusters at $z \sim 1$.  If the strong
redshift evolution in clusters predicted by CDM is correct, then the more
interesting mass regime of collapsed objects at $z > 1$ may be that of poor
clusters not yet entirely virialized.  Near--IR imaging surveys may be able
to provide a cluster sample at very high redshift complete to a relatively
low mass limit.  We describe here the discovery of one $z > 1$ cluster which
resulted from a near--IR field survey.   Except where noted, the assumed
cosmology is H$_0 = 65$ km s$^{-1}$ Mpc$^{-1}$ and q$_0 = 0.1$.

\section{Observations}

\subsection{Survey Imaging}

Elston, Eisenhardt, \& Stanford (1997, hereinafter EES) have recently
completed a $BRIzJK$ field survey designed to cover $\sim$100 arcmin$^2$ over
four areas of high galactic latitude sky down to $K \sim 22$.  The
observations were carried out at the 4~m telescope of the Kitt Peak National
Observatory.  Optical imaging was obtained with the PFCCD/T2KB, which gives
0.48$\arcsec$ pixels over a 16$\arcmin$ field. The IR imaging was obtained
with IRIM, in which a NICMOS3 HgCdTe array provides 0.6$\arcsec$ pixels over
a $\sim$2.5$\arcmin$ field.  Details of the observing and reductions are
given in EES.  The $BRJK$ survey images were coaligned and convolved to the
same effective PSF of FWHM$\approx$1.5$\arcsec$.  Calibrations of the optical
and IR images onto the Landolt and CIT systems were obtained using
observations of Landolt and UKIRT standard stars, respectively.  The $I$ and
$z$ survey data are not considered here due to problems with fringing and 
photometric calibration.

A catalog of objects within the 28 arcmin$^2$ Lynx portion of the EES survey
was obtained from the $K$ image using FOCAS (Valdes 1982), as revised by
Adelberger (1996).  The image was smoothed with a $2\farcs4 \times 2\farcs4$
kernel, and object detection was performed down to an isophotal level
corresponding to 4 $\sigma$ above sky, with a minimum object size of 3.2
arcsec$^2$.  Photometry was obtained through 2$\arcsec$ diameter apertures
using this catalog on the $BRJK$ images; the 4 $\sigma$ limit in the $K$ band
is 21.4.  A sample of very red objects was then selected using a $J-K \gtrsim
1.9$ criterion to search for $z > 1$ galaxies.  Serendipitously, a large
overdensity of these very red objects appeared in a small region in 
the Lynx survey field.  Of the 103 objects with $J-K \gtrsim 1.9$ down to $K
= 21.0$ in the entire 28 arcmin$^2$ field, 25 fall within a circular area of
radius $\lesssim 45\arcsec$.  In this region, the density of the very red
objects (hereinafter referred to as a spatially compact group, or SCG) is
13.6 arcmin$^{-2}$, in contrast to 2.8 arcmin$^{-2}$ over the rest of the 28
arcmin$^2$ of the Lynx survey field.

Additional $JHK$ imaging of a smaller area at the SCG was obtained at the
KPNO 4m telescope with IRIM in February 1997.  After standard reductions, the
new $JK$ imaging was combined with that from the field survey in the SCG area
to produce deep $JK$ images, which we use in the remainder of this paper
in place of the survey $JK$ imaging.  The new data were obtained to improve
the photometry of the objects in a 9 arcmin$^2$ area at the SCG.  The summed
images have 4$\sigma$ limits of $K = 21.7$, $H = 22.4$ and $J = 23.1$ in 2.4
arcsec diameter apertures.  Figure 1 (plate X) shows greyscale images in the
$K$ and $R$ bands of the SCG, with the $J-K \gtrsim 1.9$ objects highlighted.

\subsection{Keck Spectroscopy}

Based on the discovery of the SCG, we proceeded to obtain optical
spectroscopy in a limited area of the Lynx survey field.  Objects were
selected using the $J-K$ color and the $R$ magnitude, and are listed in Table
1.  We prepared a slitmask including slitlets for 14 objects with $J-K
\gtrsim 1.9$, 6 of which are within the SCG.  Five other objects outside the
SCG were included so as to fill out the mask.  The slits had widths of 1.5
arcsec and minimum lengths of 10 arcsec.  The slitmask was used with the Low
Resolution Imaging Spectrograph (LRIS; Oke et al.\ 1995) on the 10~m Keck II
telescope on 03-04 February 1997 UT to obtain deep spectroscopy.  We used the
400 l mm$^{-1}$ grating, which is blazed at 8500 \AA, to cover a nominal
wavelength range of 6000 to 9700 \AA, depending on the position of a slit in
the mask.  The dispersion of $\sim$1.8 \AA~pixel$^{-1}$ resulted in a
spectral resolution of $\sim$9 \AA, which is the FWHM of the emission lines
in the arc lamp spectra.  We obtained 6 $\times$ 1800~s exposures with this
setup in photometric conditions with subarcsec seeing.  Objects were shifted
along the long axis of the slits between exposures to enable better sky
subtraction.

The slitmask data were separated into 19 individual spectra and then reduced
using standard longslit techniques.  The 6 exposures for each object/slitlet
were reduced separately and then coadded for each night separately.  The
spectra were reduced both without and with a fringe correction; the former
tends to yield higher quality object spectra at the shorter wavelengths,
while the latter is necessary at the longer wavelengths.  Wavelength
calibration was obtained from arc lamp exposures taken immediately after the
object exposures.  A relative flux calibration was obtained from a longslit
observation of the standard stars HZ 44 and G191B2B with the 400 l mm$^{-1}$
grating.  While these spectra do not straightforwardly yield an absolute flux
calibration of the slit mask data, the relative calibration of the spectral
shapes should be accurate.  One--dimensional spectra were extracted for each
of the 19 objects in the 2 sets of summed images from the 2 observing nights.

\subsection{Rosat Archive}

A search of the Rosat Archive resulted in the discovery of a deep PSPC
observation which includes the SCG within the on--axis field of diameter
$\sim$40$\arcmin$.  A 64 ksec PSPC observation was obtained in two exposures
(sequence rp900009p) on two dates in 1991 centered at $\alpha = 08^h 49^m
12\fs0, \delta = +44\arcdeg 50\arcmin 24\arcsec$ (J2000), which is
$\approx$7.5 arcmin from the center of the SCG.  We obtained these exposures
from the Rosat Archive and processed them into two images using all available
channels.  The two reduced images were then summed, after a very small
spatial shift to achieve coalignment, to produce a single image of the field
in the 0.1--2.4 keV band.

\section{Results}

\subsection{Optical Spectroscopy}
Spectral features characteristic of galaxies were identified in all but 2 of
the 19 spectra; one object (\#109, the bluest in $J-K$) proved to be a star
and another (\#262, $R = 25.4$) was too faint to classify.  We calculated
redshifts where possible from well--known rest frame stellar absorption and
emission lines such as Ca II H+K and [O II]$\lambda3727$; these are given in
Table 1.  All 6 of the targeted objects in the SCG were found to lie at
approximately the same redshift, with an average value of 1.273.  Two more
galaxies at distances of $\sim$2 arcmin from the SCG center lie at $z = 1.27$
as well.  Another object, \#148, was found to have the same redshift, but
lies nearly 4 arcminutes from the SCG center.  We will assume that it is too
far away to be considered part of the apparent cluster.  A histogram of the
18 redshifts determined from our spectroscopy is shown in Figure 2.  Of the
14 objects with $J-K \gtrsim 1.9$, only \#82 has $z < 1$.  This high success
rate lends credence to the model--based prediction that a red $J-K$ color is
a reliable way to identify high redshift galaxies.

The 8 spectroscopic members are listed in Table 2, which gives the identified
spectral features in each case.  Spectra of most of the members show
absorption lines (Ca II H+K, MgI $\lambda 2852$ and $\lambda 3830$, and MgII
$\lambda 2800$) and spectral breaks (D4000, B2900, B3260 [Hamilton 1985,
Fanelli et al.\ 1990]) similar to present--epoch ellipticals.  An example,
the 3 hour summed spectrum of the galaxy brightest at $R$ (\#65), is shown in
Figure 3.  Two of the $z = 1.27$ galaxies have significant [O II] emission;
these are also relatively blue in $R-K$.  To better determine the redshifts
of the 8 candidate cluster members, we used the fourier quotient technique as
implemented in the FXCOR task of IRAF.  The 6 red members were
cross--correlated with a spectrum of M32, and the two emission--line galaxies
with an Sc template (Kinney et al.\ 1996).  The resulting redshifts, along
with their uncertainties, are listed in Table 2.  Based on these 8 members,
we adopt a nominal center for 0848+4453 at $\alpha = 8^h 48^m 34\fs6$,
$\delta = +44\arcdeg 53\arcmin 42\arcsec$ (J2000).

We have calculated an estimate of the line of sight velocity dispersion in
the candidate cluster, which should be treated with caution due to the small
number of members.  Following the recommendation of Beers, Flynn, \& Gebhardt
(1990) for a very small data set, we used the ``gapper'' method, as
implemented by their ROSTAT package, to calculate $\sigma = 740 \pm 190$ km
s$^{-1}$ in the cluster rest frame.  A somewhat more intuitive, if less
robust, estimate may be obtained from the classical standard deviation
estimator.  Assuming an underlying Gaussian for the galaxy velocities, we
find that $\sigma = 700 \pm 180$ km s$^{-1}$.  These dispersion estimates
approximately correspond to the median value for Abell richness class $R = 1$
(Zabludoff et al.\ 1993).  If we assume an isothermal gas sphere model, we
can calculate the mass contained within a given radius from the velocity
dispersion.  Using $\sigma = 700$ km s$^{-1}$, we find that the mass within
the Abell radius, $1.5h_{100}^{-1}$ Mpc, is $M = 3.5^{+2.0}_{-1.5} \times
10^{14} h_{100}^{-1}~M_{\sun}$.  We have chosen to use the Abell radius for
mass estimates because it is similar to the virial radius often used in
theoretical calculations.

\subsection{Optical--IR Photometry}
We have obtained more accurate photometry for the objects in the SCG by
making use of the more recent February 1997 imaging data set.  Object
detection (as well as the subsequent photometry) on the $K$ image of the SCG
was performed with FOCAS using parameters similar to those of the initial
catalog described in \S 2.1.  Photometry of objects in the resulting catalog
was obtained on the $BRJHK$ images in 2.4$\arcsec$ diameter apertures.  The
Galatic $E(B-V) \approx 0.04$ toward the Lynx field is small, so we have not
applied an extinction correction to our photometry.  $BRJHK$ photometry of
all 8 spectroscopic members within the SCG is given in Table 3.  Objects \#95
and 181 lie outside the area of the February 1997 imaging so only $BRJK$
magnitudes based on the survey imaging (EES) are available.  The errors in
Table 3 do not include systematics due to the uncertainties in e.g.\ the
zeropoints and PSF matching.  We estimate the quadrature sum of the
systematic errors in the photometry to be $\sim$0.05 for the IR colors, and
$\sim$0.04 for the optical-$K$ colors.

The $J-K$ and $R-K$ vs $K$ color--magnitude diagrams for all objects in a 9
arcmin$^2$ area around the SCG are shown in Figure 4.  Also plotted in both
panels of Figure 4 are estimates of the no--evolution color--magnitude locus
for early--type galaxies.  The dashed lines were calculated as described in
SED98 using photometry of early--type galaxies from a $UBVRIzJHK$ dataset
covering the central $\sim$1 Mpc of the Coma cluster (Eisenhardt et al.\
1997).  This data set enables us to determine, by interpolation, the colors
that Coma galaxies would appear to have if the cluster could be placed at $z
= 1.27$ and observed through the $RJK$ filters used on 0848+4453.  Most of
the objects within the SCG at $K < 20$ are significantly bluer than
present--epoch cluster ellipticals in $R-K$, and similar or slightly bluer in
$J-K$.  Because the uncertainty in the zeropoint of the Coma no--evolution
prediction is similar to the amount of expected color evolution in $J-K$, we
do not find strong evidence for passive evolution of the cluster galaxies in
the observed $J-K$ color.

The colors of most of the spectroscopic members (the encircled points in
Figure 4) are broadly consistent with the predictions of a ``standard'' BC97
elliptical galaxy model for $z = 1.27$.  By standard we mean the case of
passive evolution of a 1 Gyr burst stellar population with solar metallicity
formed at $z = 5$ for $h=0.65$, q$_0 = 0.1$.  Such a model predicts $R-K =
6.0$ and $J-K = 1.9$ at $z=1.27$ for a galaxy age of 3.25 Gyr.  It is
noteworthy that the same ``standard'' passive--evolution model was also found
to provide a reasonable fit to the average optical--IR colors of early--type
galaxies in a large sample of clusters at $0.3 < z < 0.9$ (SED98).  For $z_f
< 3$ in our standard BC97 model, the predicted $R-K$ becomes uncomfortably
blue with respect to the colors of the known $z=1.27$ galaxies (except for
\#237; see below).  Ages less than $\sim$3 Gyr for the stellar populations
could be accomodated if the mean metallicity were greater than solar.  Also,
we have made no attempt to correct the colors for extinction due to dust
internal to the galaxies.  If this is large, then the true colors of the
cluster galaxies would be bluer than our measured values, resulting in a
better fit to models with $z_f < 3$.

The 6 spectroscopic members in the central area of the cluster broadly fall
into two groups, the red objects (\#65, 70, 108, 135, 142), and the lone blue
object, \#237.  In Figure 5, we have plotted broad--band spectral energy
distributions (s.e.d.) of the two groups constructed from their observed
$BRJHK$ band photometry.  The red objects were normalized at $K$ and averaged
together, without weighting.  Also shown in Fig.\ 5 are two BC97 models.  The
``standard'' elliptical model described above provides a reasonably good fit
to the average s.e.d.\ of the red objects, except at rest frame
$\sim$1900$\AA$.  To better fit \#237, a second model is shown, which combines
a 1 Gyr burst with a 1\% (by mass) burst added at an age of 2.25 Gyr, $i.e.$
the recent starburst has an age of 1 Gyr at the 3.25 Gyr age of the composite
model.  In both cases, the models are scaled so as to match the flux
densities at the reddest (observed $K$) band.

Using our standard BC97 elliptical model and our assumed cosmology, we
calculated the absolute magnitudes in the rest frame $V$ band from the
observed frame $J$ for the cluster members; these are listed in Table 3.  The
$M_V$ range from $-23.5$, about 2 magnitudes brighter than $L^\ast$ in
present--epoch clusters (Colless 1989), to $-21.7$ for the bluest galaxy.
The BC97 model predicts $\approx$1.3 magnitudes of luminosity evolution in
$M_V$ from $z=1.27$ to $z=0$ in our assumed cosmology, suggesting that the
most luminous of the cluster members would evolve into $\sim$2$\times L^\ast$
ellipticals in the present epoch.

\subsection{X-ray Imaging}
The summed PSPC image yields a detection of 89 $\pm$ 19 counts in a 60 arcsec
radius region centered on the nominal center of the spectroscopically
identified members.  The centroid of the X-ray counts in this region lies at
$\alpha = 8^h 48^m 35\fs1$, $\delta = +44\arcdeg 53\arcmin 49\arcsec$
(J2000), which is within $\sim9\arcsec$ of the nominal optical cluster
center. The flux in the observed 0.1--2.4 keV band is $1.8 \times 10^{-14}$
ergs s$^{-1}$ cm$^{-2}$ in a 2 arcmin diameter region at the cluster.  The
origin of the X--ray flux could be several types of sources, such as
lower--redshift galaxy groups, AGN within the SCG, or hot gas in the apparent
cluster at $z=1.27$.  As seen in Figure 2, our limited optical spectroscopy
does not show any significant groups along the line of sight towards the
$z=1.27$ galaxies, although our selection criteria for the spectroscopic
observations introduces some bias into the redshift distribution.  No object
in the cluster field is detected down to $\sim$20 mJy at 327 MHz in the
Westerbork Northern Sky Survey (WENSS; de Breuck, personal communication).
This suggests there are no active galaxies in the SCG which could account for
the observed X--ray emission.  However, it is possible that an X--ray loud
but radio--quiet AGN could have escaped detection by WENSS, and in fact
several objects with blue $B-R$ colors consistent with the properties of AGN
lie within the SCG.  But the area of interest is very small, so the
likelihood of there being an AGN is very small.  According to Boyle et al.\
(1993), we should expect to find only $\sim$0.02 X-ray AGN in the $2\farcm0$
region of the PSPC detection with $S_{\nu} = 1.8 \times 10^{-14}$ ergs
s$^{-1}$ cm$^{-2}$.

The third possibility is that the PSPC detection arises from hot gas trapped
in a deep potential well at $z=1.27$.  Assuming a gas temperature of 6 keV
and a Galactic neutral hydrogen column density of N(H) $= 2.7 \times 10^{20}$
atoms cm$^{-2}$, the luminosity in the rest frame 0.1--2.4 keV band is $L_x =
1.5 \times 10^{44}$ ergs s$^{-1}$ for $h_{100} = 0.65$ and q$_0 = 0.1$.  Although
the procedure is very uncertain, we can use this $L_x$ to estimate the
cluster mass.  We must assume that the gas is isothermal, that the total mass
is distributed like the X-ray emitting gas, and that the gas temperature
$T_{keV}$ is related to $L_x$ by the relationship determined at low-$z$ by
Edge \& Stewart (1991).  Then, following Donahue et al.\ (1997), the total
mass within a given radius $r$ is $M(<r) \sim 1.1 \times 10^{14}~M_{\sun}~
\beta~T_{keV}~R_{Mpc}$.  Assuming $\beta = 0.8$, we find that the total mass
within the Abell radius $r = 1.5 h_{100}^{-1}$ Mpc is $5.1 \times
10^{14}h_{100}^{-1}~M_{\sun}$.

The apparent confirmation of the cluster by the detection of X--ray emission
from the SCG is qualified by the fact that the PSPC PSF is too broad to
resolve the X--ray emission.  Figure 6 shows a comparison of the growth curve
of the detected X--ray source with that of a nearby star.  The broader growth
curve of the former is consistent with a cluster origin, but the difference
is insufficient to rule out the possibility that the observed X--ray flux is due
to a single galaxy, either in the apparent cluster, or along the line of
sight.  Nevertheless, the existence of a spatially--coincident X--ray source
of a luminosity reasonable for a moderate--size cluster at $z=1.27$ strongly
supports the identification of the SCG as a gravitationally--bound system.

\section{Discussion}

The observations presented here provide strong evidence supporting the
identification of a very red SCG found in a near--IR field survey as a galaxy
cluster at $z = 1.27$.  The density of $J-K \gtrsim 1.9$ objects in the SCG
is 5 times higher than in the rest of our Lynx survey field.  The spectroscopy
shows that 6 galaxies are at the same redshift within the SCG, and another 2
more radially--distant objects are also at $z=1.27$.  Finally, an archival
Rosat PSPC observation detects X--ray emission coincident with the SCG.  The
fact that no AGN are known to exist in the SCG suggests that the X--ray
emission is most likely due to hot gas in the potential well of a galaxy
cluster.  We conclude that 0848+4453 is currently the highest redshift
cluster discovered without targetting a central active galaxy.

\subsection{Member Galaxies}
Given that we have a $z=1.27$ galaxy cluster, the evolutionary state of the
stellar populations in the member galaxies at this substantial lookback time
is of great interest.  SED98 found that passively--evolving elliptical
galaxies, formed at high redshift, inhabit cluster cores back to at least
$z=0.9$.  The $R-K$ colors of many of the objects in the cluster area show
$\sim$1 mag of color evolution relative to our Coma--based no--evolution
prediction for early--type galaxies.  The $RJK$ colors generally agree with
the predictions of a passively--evolving elliptical model at $z=1.27$ for
$z_f > 3$ in a $h_{100}=0.65$, q$_0 = 0.1$ cosmology.  The amount of passive
evolution in $R-K$ is in good agreement with that found by SED98.  The
relatively blue $B-K$ color displayed in the red objects' average s.e.d.\ in
Figure 5 indicates there is some uncertainty in the identification of the red
members as being elliptical galaxies.  But Fig.\ 5 also shows that it takes
only a very small amount of recent star formation to explain the excess flux
at rest frame $\sim$1900 \AA.  Most of the members in the cluster have
optical spectra with the breaks and absorption lines of the evolved stellar
populations seen in elliptical galaxies.  The D4000 is clearly evident in 5
of the spectroscopic members, as well as B3260 and B2900 in some.  A more
detailed spectral analysis, akin to that in Spinrad et al.\ (1997)
of an elliptical at $z=1.55$, is deferred to a later paper.

While it is tempting to try constraining the formation epoch of the galaxies,
such age estimates are very uncertain (Charlot, Worthey, \& Bressan 1996),
due in part to the degeneracy of age with metallicity.  Also, it is difficult
to distinguish between the formation of the stellar populations and the
structural formation of the galaxies.  Two of the cluster members, \#237 and
181, are considerably bluer in $R-K$ and $B-K$ than the rest, and also
exhibit [O II] emission.  Deciding if these properties mean that these
objects are late--type galaxies, or early--types with some recent star
formation, is difficult without more detailed morphological information.

\subsection{Cluster Properties}
Given its high redshift, the mass of the cluster is of particular interest.
Neither gravitational arcs nor weak lensing has been observed in
ground--based imaging, so we cannot make use of such mass estimators.  Both
the $L_x$ and the $\sigma$ of the member galaxies indicate that 0848+4453
falls toward the low end in the mass range of low--redshift clusters (Ebeling
et al.\ 1997; Nichol et al.\ 1997; Zabludoff et al.\ 1993).  At high
redshift, knowledge of the cluster distribution functions in velocity
dispersion and X--ray luminosity is much poorer.  Also, it is unclear if the
relationships between velocity dispersion and mass, and X--ray luminosity and
mass remain the same as in the present epoch.  The correlation between $L_x$
and $\sigma$ may be shifted towards higher $\sigma$ for a given $L_x$,
relative to the relation at low redshift (Bower et al.\ 1997).  In the
context of CDM, one can imagine that calculations of $\sigma$ are
overestimated as a result of bound but not yet virialized galaxies,
as has been suggested by the comparison of lensing mass estimates
with $L_x$ and $\sigma$ in rich clusters at $z \sim 0.4$ by Smail et al.\
(1997).  The $L_x$ and $\sigma$ of 0848+4453 follow the low-$z$ relationship
between cluster velocity dispersion and X--ray luminosity (Edge \& Stewart
1991b), which may mean that the known member galaxies are virialized.  On the
other hand, the fact that most of the members appear to be early--type
galaxies could bias the calculated dispersion.  Using the CNOC sample of
moderate--redshift clusters, Carlberg et al.\ (1997) found that the velocity
dispersion of the blue members was about 30\% greater than that of the red
cluster galaxies.  So our estimate of the dispersion may be somewhat lower
than the true value for all cluster members.  Also, our $L_x$-based mass
estimate is very uncertain, given that we know very little about the physical
state of the X--ray emitting gas.

If our mass estimates for 0848+4453 are approximately correct, then it is
interesting to consider the resulting constraint on theories of structure and
$\Omega_0$.  The predictions of Eke, Cole, \& Frenk (1996) for cluster
evolution in an $\Omega_0 = 1$ universe indicate that we should not expect to
find any massive ($M > 3.5 \times 10^{14} h_{100}^{-1}~M_{\sun}$ according to
Eke et al.) clusters at $z > 0.5$.  The existence of 0848+4453, as well as
other truly massive clusters at high redshift such as MS 1054.5-0321 (Luppino
\& Kaiser 1997; Donahue et al.\ 1997) at $z = 0.828$, 3C 184 at $z=0.996$
(Deltorn et al.\ 1997), and 3C~324 at $z = 1.206$ (Dickinson 1996), indicate
that $\Omega_0 < 1$ and/or CDM is incorrect.  The predicted number density of
massive clusters at high-$z$ for an open universe is considerably higher
(Eke, Cole, \& Frenk 1996).  For $\Omega_0 = 0.3$, $\Lambda_0 = 0$ there
should be only 0.01 massive clusters at $z < 1.3$ in the 100 arcmin$^2$ of
the EES survey.  If its mass is in fact considerably lower than our
estimates based on the velocity dispersion and the $L_x$, then 0848+4453
could fit qualitatively, if not quantitatively, in the scenario envisioned by
CDM for the growth of structure in a low $\Omega$ universe.  In this case, if
we could follow it through cosmic time, we would expect to see 0848+4453 grow
through mergers with other groups to become a massive cluster by the present
epoch.  Even more speculatively, if we were to trace 0848+4453 back to higher
redshift, we might expect to see a structure similar to the concentration of
15 Lyman break galaxies at $z = 3.1$, located within a 11$\arcmin$ by 8$\arcmin$
area, which were found within a 9$\arcmin$ by 18$\arcmin$ survey field by Steidel
et al.\ (1997).

The discovery of 0848+4453 was based on the expectation that the $J-K$ color
is a reasonable redshift indicator (out to $z \sim 2$) for nearly all galaxy
types.  This expectation has been confirmed by our finding that of the 14
galaxies with $J-K \gtrsim 1.9$ for which spectra were obtained, 13 do indeed
have $z > 1$.  In the not too distant future near--IR imaging surveys may be
able to provide large samples of $z > 1$ clusters covering much of the
cluster mass range in existence at that epoch.  Such a database would prove
invaluable in both the study of large--scale structure and of galaxy
evolution.

\acknowledgments

We thank Ed Moran for processing the Rosat PSPC data, and Drew Phillips for
the use of his slitmask software.  RE and PRE thank Kitt Peak National
Observatory for the support and telescope time provided to their survey.  The
work by SAS at LLNL was performed under the auspices of the US Department of
Energy under contract W-7405-ENG-48.  Extragalactic research by HS and DS is
supported by NSF grant 95-28536.  Portions of the research described here
were carried out at the Jet Propulsion Laboratory, California Institute of
Technology, under a contract with NASA.  The W.\ M.\ Keck Observatory is a
scientific partnership between the University of California and the
California Institute of Technology, made possible by a generous gift of the
W.\ M.\ Keck Foundation.

\newpage

\begin{figure}
\epsscale{0.6}
\plotone{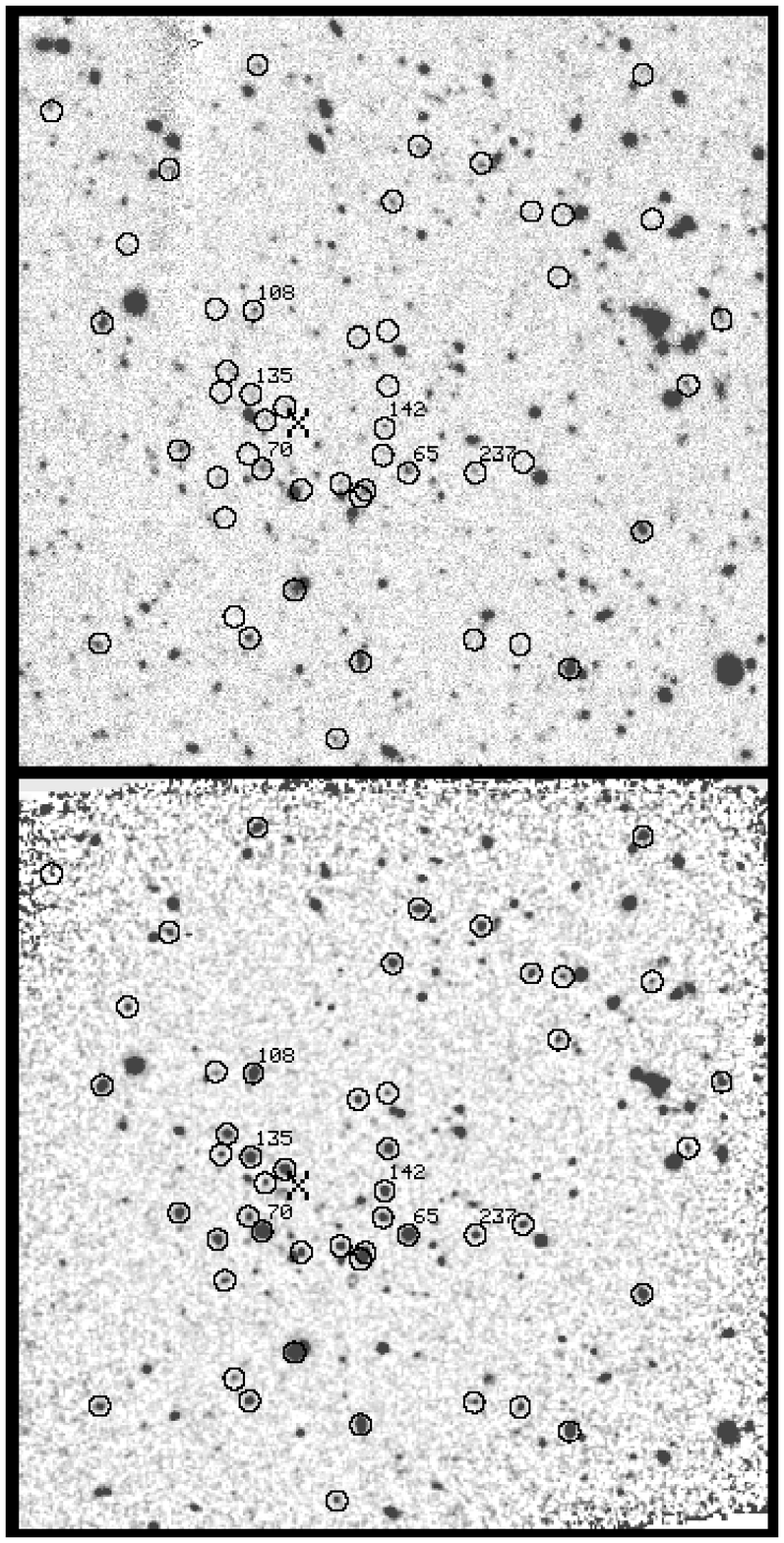}
\caption{$K$ (bottom) and $R$ (top) band images of a $3\farcm2 \times 3\farcm2$
area at the SCG.  Objects with $J-K \gtrsim 1.9$ are marked by the
circles, and the 6 spectroscopically confirmed member galaxies within the SCG
are marked with their ID numbers to the upper right.  A cross marks the
centroid of the PSPC detection. North is up and East is left in the images,
which are centered at $\alpha = 8^h 48^m 32\fs7$, $\delta = +44\arcdeg
53\arcmin 56\arcsec$ (J2000).}  \epsscale{1.0}
\end{figure}

\begin{figure}

\plotone{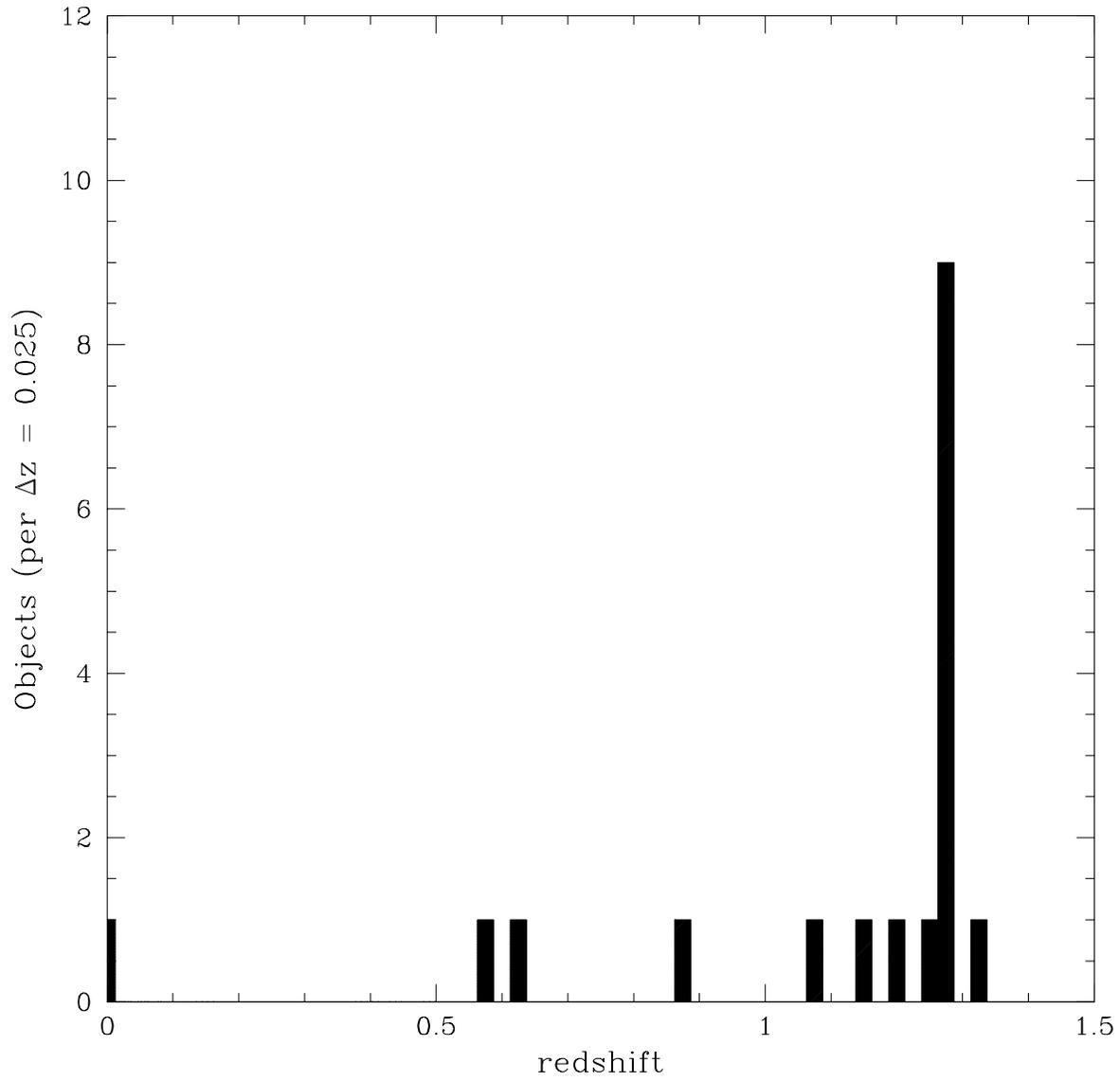}
\caption{A histogram of the 18 redshifts determined in the Keck/LRIS slitmask observation.}

\end{figure}

\begin{figure}
\epsscale{0.9}
\plotone{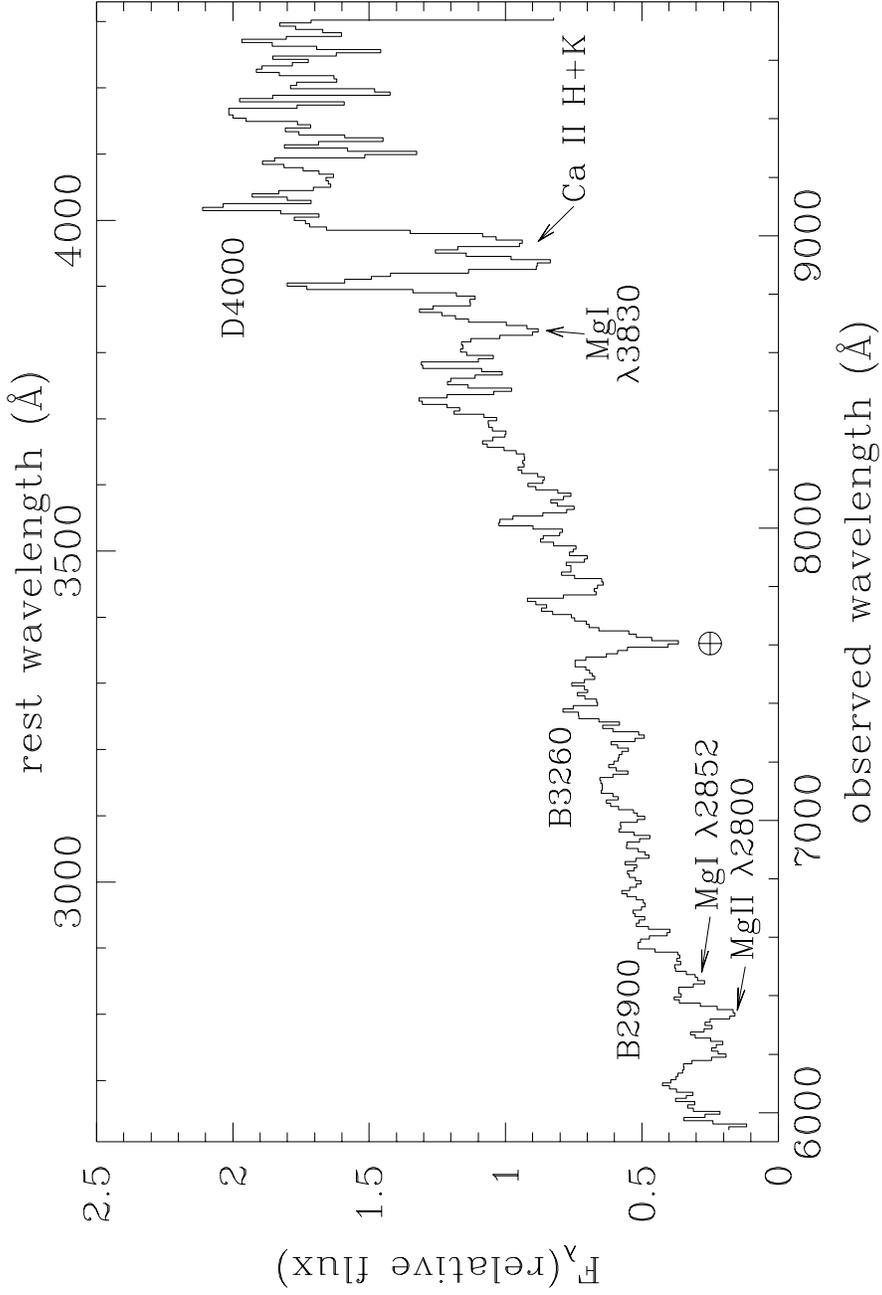}
\caption{Optical spectrum of Object 65 ($z=1.264$, $R = 24.0$) in ClG
0848+4453 after being smoothed by an 11 pixel boxcar.}
\epsscale{1.0}
\end{figure}

\begin{figure}

\plotone{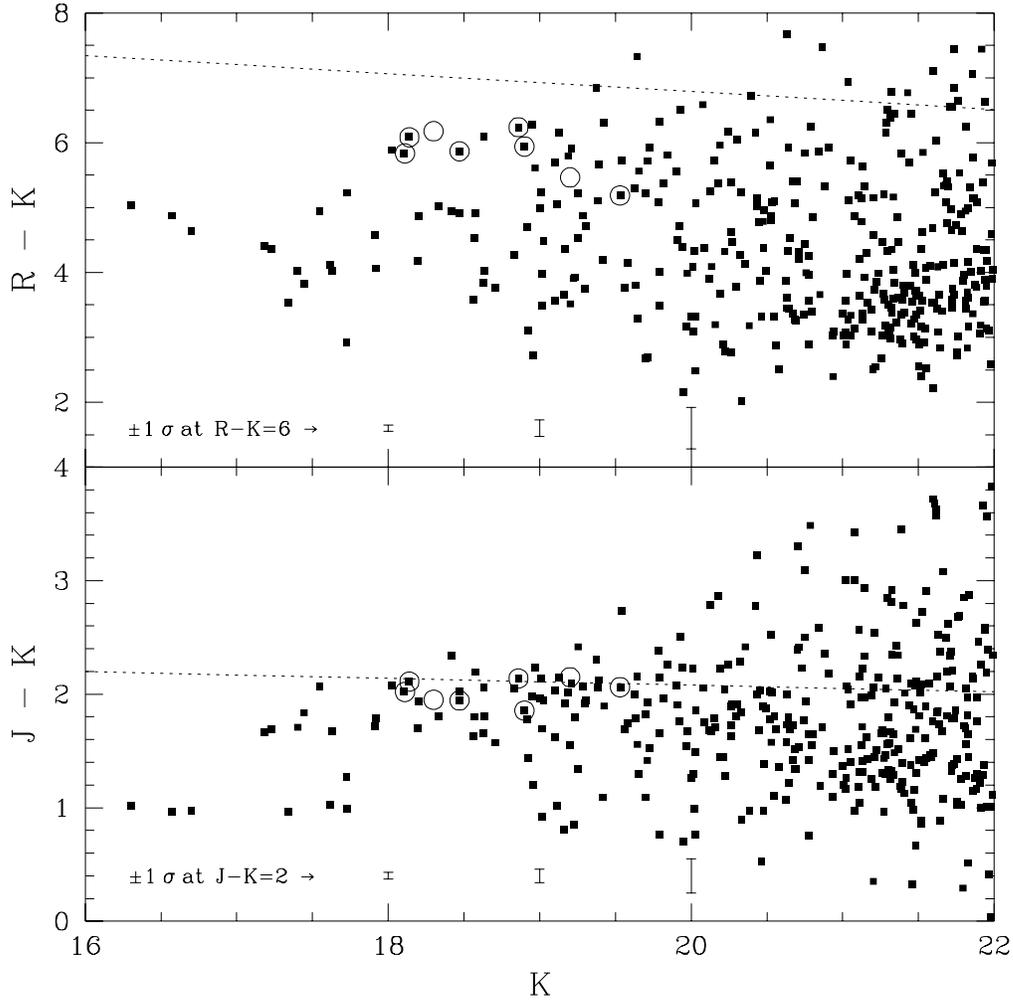}
\caption{Color--magnitude diagrams of the 9 arcmin$^2$ area at the SCG.  The
spectroscopically identified galaxies at $z=1.27$ are marked with circles;
note that two of these lie outside of the 9 arcmin$^2$ area (but do fall
within the larger survey field) so do not appear with black squares
inside. The dashed lines represent the no--evolution prediction for
elliptical galaxies, as described in the text.  The errors in the colors at
$J-K = 2$ and $R-K = 6$ are shown at the bottom of the panels. }

\end{figure}

\begin{figure}
\epsscale{0.85}
\plotone{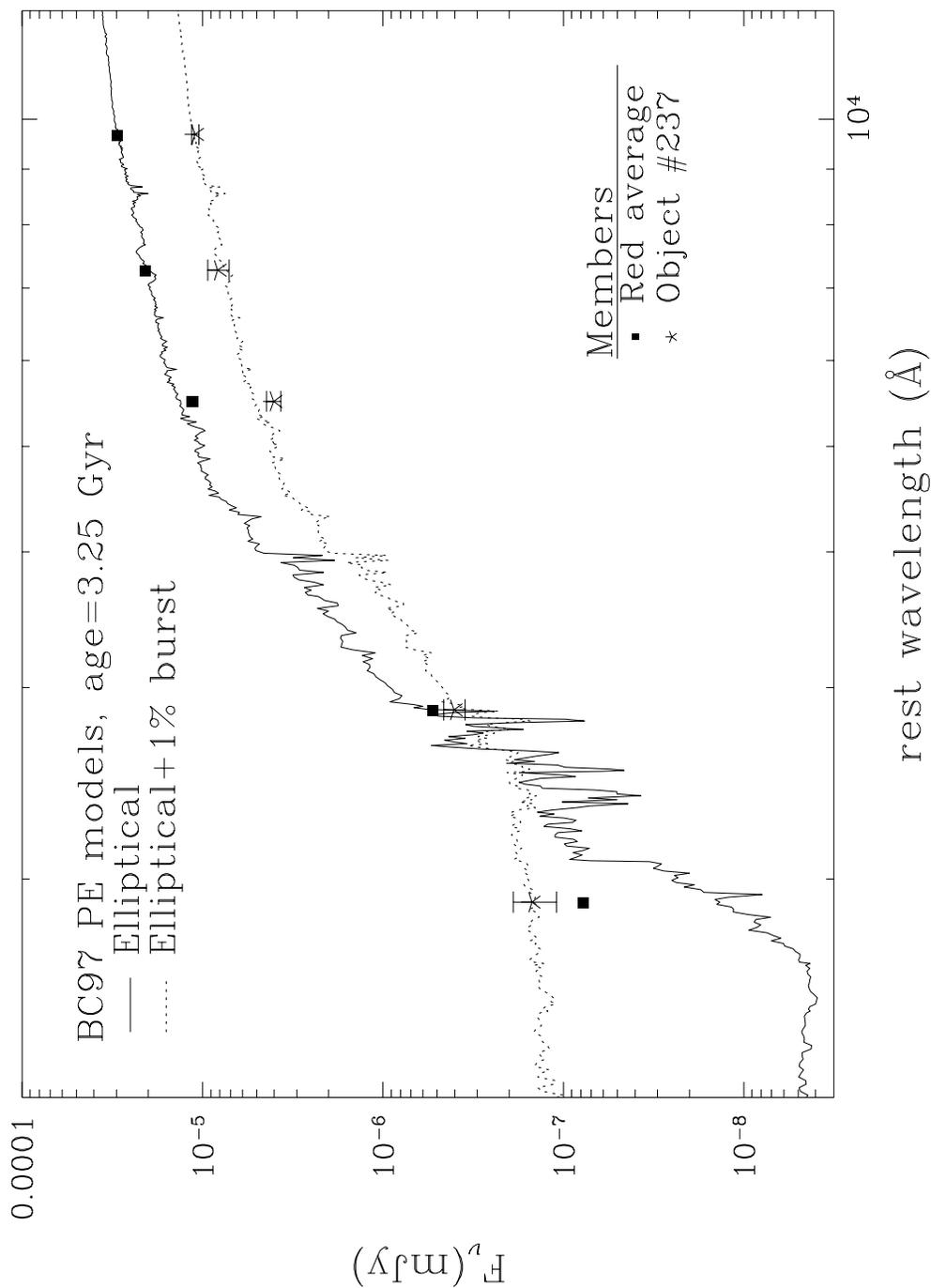}
\caption{Rest frame spectral energy distributions of the spectroscopic
members.  The solid squares represent the average of the 5 red galaxies within
the SCG, and the open stars (with $\pm$1 $\sigma$ errorbars) represent 
\#237.  The two spectra plotted in the figure were generated from BC97 passive
evolution models, and are described fully in the text.}
\epsscale{1.0}
\end{figure}

\begin{figure}

\plotone{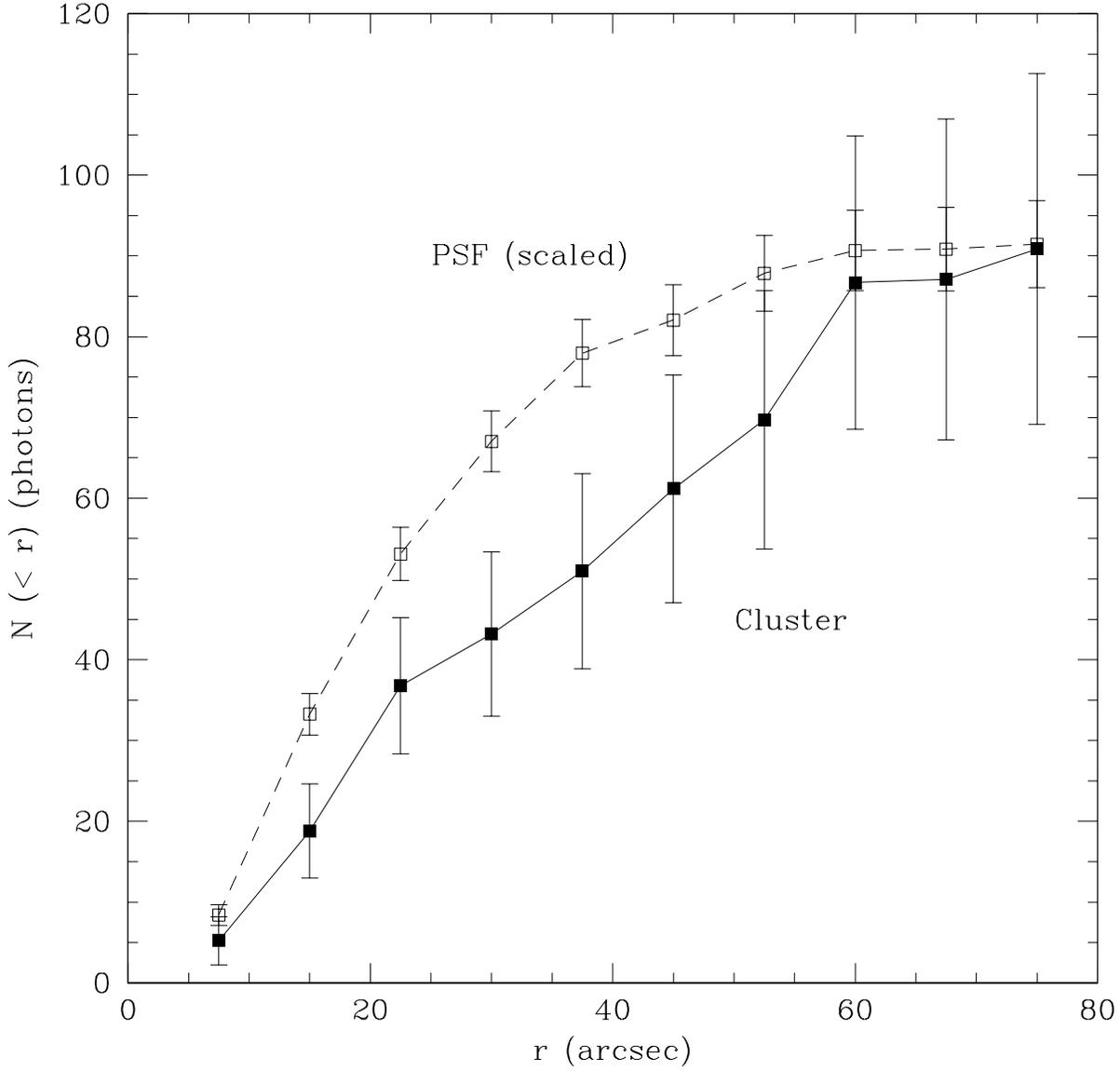}
\caption{The Rosat/PSPC growth curve of a nearby star compared to that of the
X--ray source at 0848+4453.  $\pm$1 $\sigma$ error bars are shown.  The
nominal FWHM of PSPC data within the on--axis field is $\approx$23 arcsec;
the cluster is unresolved within the errors.}

\end{figure}

\newpage
\hoffset -0.25in
\begin{deluxetable}{ccccccc}
\small
\tablenum{1}
\tablecaption{Summary of Spectroscopic Targets in Lynx Field}
\tablehead{
\colhead{ID} & \colhead{R.A.} & \colhead{Dec.} & \colhead{$z$} &
\colhead{$J-K$} & \colhead{$R$}  & 
\colhead{Radius\tablenotemark{a}} 
}
\startdata

237 & 08:48:30:73 & 44:53:36 &1.271 &2.1 &24.7 &37\nl
65  & 08:48:32.36 & 44:53:36 &1.263 &2.0 &24.0 &19\nl
142 & 08:48:32.92 & 44:53:47 &1.277 &1.9 &24.8 &18 \nl
70  & 08:48:35.93 & 44:53:37 &1.275 &2.1 &24.2 &20 \nl
135 & 08:48:36.18 & 44:53:56 &1.276 &2.1 &25.1 &22 \nl
108 & 08:48:36.06 & 44:54:18 &1.277 &1.9 &24.4 &45\nl
95  & 08:48:32.51 & 44:51:41 &1.268 &1.9 &24.4 &129\nl
181 & 08:48:45.54 & 44:54:31 &1.270 &2.1 &24.7 &136\nl
207 & 08:48:43.97 & 44:53:11 &1.065 &1.9 &25.0 &107 \nl
262 & 08:48:41.98 & 44:55:02 &\nodata &1.3 &25.4 &122 \nl
167 & 08:48:46.80 & 44:55:01 &1.24\phn &1.9 &24.3 &162 \nl
200 & 08:48:53.24 & 44:53:49 &0.875 &1.7 &23.9 &277\nl
162 & 08:48:50.21 & 44:55:10 &1.138 &1.9 &24.5 &199 \nl
118 & 08:48:50.84 & 44:55:34 &1.328 &1.9 &24.5 &217 \nl
148 & 08:48:52.92 & 44:55:21 &1.273 &1.8 &24.3 &230\nl
109 & 08:48:52.65 & 44:56:13 &0\tablenotemark{b}&0.8 &23.3 &206 \nl
305 & 08:48:56.28 & 44:55:51 &1.21\phn&1.9&24.9 &276 \nl
82  & 08:48:57.56 & 44:56:11 &0.622 &1.9 &23.1 &299 \nl
55  & 08:48:57.06 & 44:56:48 &0.569 &1.6 &22.8 &310 \nl

\enddata
\tablenotetext{a}{Radius is the distance in arcsec from the SCG
center}
\tablenotetext{b}{M star}
\tablecomments{Coordinates are J2000}
\end{deluxetable}

\newpage
\hoffset -0.25in
\begin{deluxetable}{llcl}
\small
\tablenum{2}
\tablecaption{Spectroscopic Results for Members in ClG J0848+4453}
\tablehead{
\colhead{ID} & \colhead{$z$} & \colhead{$\delta z$} & \colhead{Major Spectral Features} 
}
\startdata
65  &1.2637 & 0.0003 &H+K, MgII $\lambda$2800, D4000, B3260, B2900 \nl
70  &1.2751 & 0.0004 &H+K, MgII $\lambda$2800, D4000, B2900, wk.\ [O II] \nl
108 &1.2783 & 0.0003 &H+K, MgII $\lambda$2800, D4000, B2900, B2640, wk.\ [O II] \nl
95  &1.2678 & 0.0005 &H+K, D4000, G-band \nl
135 &1.2758 & 0.0006 &H+K, D4000 \nl
142 &1.2789 & 0.0003 &H+K, MgII $\lambda$2800, D4000, B2900, B2640 \nl
181 &1.2706 & 0.0002 &[O II], MgII $\lambda$2800, D4000, H$\delta$, G-band\nl
237 &1.2714 & 0.0002 &[O II], MgII $\lambda$2800, H$\delta$, H+K \nl
\enddata

\end{deluxetable}

\newpage
\hoffset -0.25in
\begin{deluxetable}{lcccccc}
\small
\tablenum{3}
\tablecaption{Photometric Properties of Members in ClG J0848+4453}
\tablehead{
\colhead{ID} & \colhead{$K$} & \colhead{$H-K$} & \colhead{$J-K$} &
\colhead{$R-K$} & \colhead{$B-K$} & \colhead{M$_V$\tablenotemark{a}}  
}
\startdata
65  &$18.11\pm 0.02$&$0.81\pm 0.03$&$2.02\pm 0.03$&$5.86\pm 0.05$&$8.6\pm 0.2$&$-23.4$\nl
70  &$18.14\pm 0.02$&$1.00\pm 0.03$&$2.11\pm 0.03$&$6.09\pm 0.05$&$8.7\pm 0.2$&$-23.5$\nl
95  &$18.30\pm 0.02$&    \nodata   &$1.95\pm 0.04$&$6.18\pm 0.07$&$8.7\pm 0.2$&$-23.0$\nl
108 &$18.47\pm 0.02$&$0.89\pm 0.04$&$1.94\pm 0.04$&$5.87\pm 0.06$&$8.6\pm 0.2$&$-23.1$\nl
135 &$18.86\pm 0.03$&$1.09\pm 0.05$&$2.13\pm 0.05$&$6.24\pm 0.11$&$8.5\pm 0.3$&$-22.5$\nl
142 &$18.90\pm 0.03$&$0.84\pm 0.05$&$1.87\pm 0.06$&$5.94\pm 0.09$&$8.5\pm 0.3$&$-22.4$\nl
181 &$19.20\pm 0.05$&    \nodata   &$2.15\pm 0.10$&$5.47\pm 0.09$&$7.4\pm 0.2$&$-21.9$\nl
237 &$19.53\pm 0.06$&$0.83\pm 0.09$&$2.06\pm 0.12$&$5.19\pm 0.10$&$6.8\pm 0.1$&$-21.7$\nl
\enddata
\tablenotetext{a}{Calculation of M$_V$ assumed $h_{100} = 0.65$, q$_0 = 0.1$; see
text for details.}
\end{deluxetable}

\end{document}